\RequirePackage{fix-cm}
\documentclass[twocolumn,epjc3]{svjour3}  
\smartqed  
\RequirePackage{graphicx}
\usepackage{amsmath,amssymb,url}
\usepackage{color}

\newcommand{\feynslash}[1]{/\hspace*{-2mm} #1}
\def\be{\begin{eqnarray} &&}
\def\nonu{\nonumber \\ &&}

\def\ee{\end{eqnarray}}

\journalname{Eur. Phys. J. C}
\begin{document}

\title{Exploring the Quark-Gluon Vertex with Slavnov-Taylor Identities and Lattice Simulations }


\author{Orlando Oliveira\thanksref{e1,addr1,addr2} 
        \and
        T.   Frederico\thanksref{e2,addr2} 
        \and
        W. de Paula\thanksref{e3,addr2} 
        \and
        J. P. B. C de Melo\thanksref{e4,addr3} 
}

\thankstext{e1}{e-mail: orlando@uc.pt}
\thankstext{e2}{e-mail: tobias@ita.br}
\thankstext{e3}{e-mail: wayne@ita.br}
\thankstext{e4}{e-mail: joao.mello@cruzeirodosul.edu.br}


\institute{Centro de F\'{i}sica da Universidade de Coimbra, Departamento de F\'{i}sica, Universidade de Coimbra, 3004-516 Coimbra, Portugal \label{addr1}
           \and
           Departamento de F\'{\i}sica, Instituto Tecnol\'ogico de Aeron\'autica, DCTA, 12228-900, S. Jos\'e dos Campos, Brazil \label{addr2}
           \and
           Laborat\'orio de F\'{\i}sica Te\'orica e Computacional, Universidade Cruzeiro do Sul, 01506-000, S\~ao Paulo, SP, Brazil \label{addr3}
}

\date{Received: date / Accepted: date}

\maketitle

\begin{abstract}
The soft gluon limit of the  longitudinal part of the quark-gluon vertex is studied by resorting to non-perturbative approaches to 
Quantum Chromodynamics (QCD).  Based on a Slavnov-Taylor identity (STI), the longitudinal form factors is expressed in terms of the 
quark-ghost kernel, the quark self energy and the quark wave function. An exact  relation between the non-vanishing longitudinal
form factors is derived  for the soft gluon limit and explored to understand the behaviour of the vertex.
Within a Ball-Chiu vertex, the form factor $\lambda_1$ was analysed using recent lattice simulations for full QCD for the soft
gluon limit. The lattice data shows that the gluon propagator resumes the momentum dependence of such component of the vertex.
This connection is understood via a fully dressed one-loop  Bethe-Salpeter equation. The behaviour of the remaining
longitudinal form factors $\lambda_2(p^2)$ and $\lambda_3(p^2)$ is investigated combining both the information
of lattice simulations and the derived relations based on the STI.
\keywords{Quark-Gluon Vertex \and QCD \and Lattice QCD}
\end{abstract}

\section{Introduction and Motivation}

The quark-gluon vertex is at the heart of all hadron phenomena. 
Quark confinement and dynamical chiral symmetry breaking, two open problems in modern strong interactions, require 
certainly a good understanding of this one-particle irreducible function.  
Despite all efforts towards a first principle calculation of the quark-gluon vertex, a complete solution 
is still lacking.

The information on the structure of the vertex is crucial also for the continuous approaches to QCD 
that, typically, rely on modelling the functional dependence of the form factors associated to its various
tensor structures, assume the dominance of one of the form factors, e.g., $\lambda_1(p^2)$ and explore various truncations 
of the corresponding Dyson-Schwinger and Bethe-Salpeter kernels. 

Progress in computing the quark-gluon vertex and the fundamental QCD kernels has been slow and it
revealed quite a difficult problem
\textit{per se}~\cite{Maris99a,Alkofer:2008tt,Rojas:2013tza,CloPPNP14,Aguilar:2014lha,Williams:2014iea,Binosi:2014aea,Williams:2015cvx,Binosi:2016rxz,El-Bennich:2016qmb,Binosi:2016wcx,Aguilar:2016lbe,Jia:2017niz,Bermudez:2017bpx,Cyrol:2017ewj,Gao:2017tkg}.
In this perspective, gathering information and combining various non-perturbative approaches can be 
useful to learn more about this fundamental QCD quantity. 
Efforts along this line has already been pursued in the Landau gauge in \cite{Rojas:2013tza}, where lattice results for the gluon, ghost and 
quark propagator have been used together with the Slavnov-Taylor Identity (STI) for the quark-gluon vertex to solve
the quark gap equation that relates all these quantities and implicitly defines a coupled set of integral equations to be solved for the unknown form factors 
of the quark-ghost kernel. We remind that the STI allows to express the quark-gluon vertex in terms of the quark-ghost kernel form factors. 
The solution for the vertex obtained from the gap equation relied on a particular simplification of the full structure tensor of the quark-ghost kernel and, therefore, of the
vertex.
A similar approach to the gluon-ghost vertex can be found in~\cite{Dudal:2012zx}.

In this work, we aim to go further on the above delineated approach studying the soft gluon limit of the quark-gluon 
vertex provided by full QCD lattice simulations for $N_f = 2$ in combination with the information that the STI adds on. 
From the STI, that constraints the longitudinal part of the quark-gluon vertex, an exact relation between 
$\lambda_1$, $\lambda_2$ and $\lambda_3$ is obtained in the soft gluon limit that links these form factors and those associated 
with the quark-ghost kernel. 

Furthermore, relying on the soft gluon limit of the STI for $\lambda_1$, we explore the results from the recent Lattice QCD simulation~\cite{Oliveira:2016muq}.
In particular, we find an empirical connection linking the gluon propagator $\Delta (p^2)$ and the soft gluon limit for $\lambda_1 (p^2)$ that writes this
form factor, for momenta up to $p \sim 10$ GeV, in terms of $\Delta (p^2)$. 
As discussed, this connection can be understood in terms of a dressed one-loop  Bethe-Salpeter equation for the vertex function.
Moreover, the information coming from the STI and the lattice data for $\lambda_1$, allows the understanding of the qualitative behaviour of the remaining non-vanishing
longitudinal form factors $\lambda_2$ and $\lambda_3$, in the soft gluon limit.

\section{Notation and Definitions}

\begin{figure}[t] 
   \centering
   \includegraphics[scale=0.75]{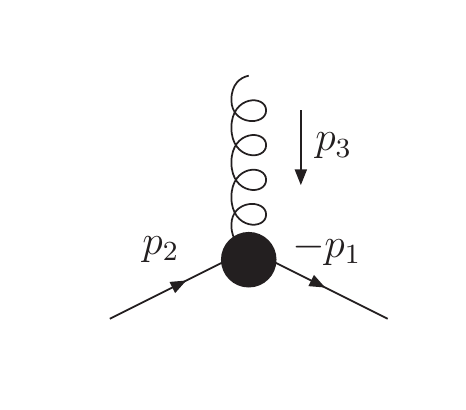} 
    \vspace*{-4mm}  
   \caption{The quark-gluon vertex for a general kinematical configuration. All momenta are incoming and verify $p_1+p_2+p_3 = 0$.}
   \label{FIG:quark_gluon_vertex}
\end{figure}

For the diagonal metric $g = ( 1, \, -1, \, -1, \, -1)$ defined in Minkowsky space,
the one-particle irreducible Green function associated to the quark-gluon vertex represented on Fig.~\ref{FIG:quark_gluon_vertex} reads
\begin{equation}
  \Gamma^a_\mu (p_1, p_2, p_3) = g \, t^a \, \Gamma_\mu (p_1, p_2, p_3) \ ,
  \label{Eq:FullVertex}
\end{equation}
where all momenta are incoming and, therefore, $p_1+p_2+p_3 = 0$, $g$ is the strong coupling constant and 
$t^a$ are the SU(3) generators in the fundamental representation. Herein we follow the notation of~\cite{Davydychev}. 

The quark propagator is color diagonal and has the following Dirac structure
\begin{equation}
 S(p) =  \frac{i}{ A(p^2) \feynslash  p\ -  B(p^2) } 
 = i \, Z(p^2) \, \frac{ \feynslash  p\ + M(p^2) }{ p^2 - M^2(p^2)},
\end{equation}
where $Z(p^2) = 1/A(p^2)$ is the quark wave function and $M(p^2) = B(p^2) / A(p^2)$ is the renormalization group invariant running
quark mass.
In the Landau gauge, the gluon propagator is given by
\begin{equation}
   \Delta^{ab}_{\mu\nu} (q) =  -i\,  \delta^{ab} \left( g_{\mu\nu} - \frac{q_\mu q_\nu}{q^2} \right) \Delta ( q^2 ) \ .
   \label{gluondress}
\end{equation}   
In the following, we will refer either to $\Delta^{ab}_{\mu\nu} (q)$ or to the form factor 
$\Delta (q^2)$ as the gluon propagator.

The Lorentz structure of the quark-gluon vertex $\Gamma_\mu$, see Eq. (\ref{Eq:FullVertex}), can be decomposed into 
longitudinal $\Gamma^{(L)}$ and transverse $\Gamma^{(T)}$ components relative to the gluon momenta as
\begin{equation}
\Gamma_{\mu}(p_1, \, p_2, \, p_3)  = \Gamma^{(L)}_{\mu}(p_1, \, p_2, \, p_3) + \Gamma^{(T)}_{\mu}(p_1, \, p_2, \, p_3),
\label{Eq_vertex}
\end{equation}
where, by definition,  
\begin{equation}
p_{3}^{\mu} ~ \Gamma^{(T)}_{\mu}(p_1, \, p_2, \, p_3)= 0 \ .
\end{equation}
By choosing a suitable tensor basis in the spinor-Lorentz space, $\Gamma_\mu$ can be written as a sum of scalar form factors that
multiply each of the elements of the basis. The full vertex $\Gamma_\mu$ requires twelve form factors and for
the Ball and Chiu~\cite{Ball:1980ay} basis one writes
\begin{eqnarray}
   \Gamma^\mathrm{L}_\mu (p_1, p_2, p_3) & = & -i\, \sum^4_{i=1} \lambda_i (p_1, p_2, p_3) ~ L^{(i)}_\mu (p_1 , p_2) \, \\ 
   \label{EQ:Lvertex} 
   \Gamma^\mathrm{T}_\mu (p_1, p_2, p_3)  & = & -i\, \sum^8_{i=1} \tau_i (p_1, p_2, p_3) ~ T^{(i)}_\mu (p_1 , p_2) \ .
   \label{EQ:Tvertex} 
\end{eqnarray}
For the longitudinal vertex studied below, the operators are given by
\begin{eqnarray}
 L^{(1)}_\mu (p_1 , p_2)  & =  &  \gamma_\mu \ ,  \\
 L^{(2)}_\mu (p_1 , p_2)  & =  & ( \feynslash p_1 - \feynslash p_2 ) \left( p_1 - p_2 \right)_\mu \ ,   \\
 L^{(3)}_\mu (p_1 , p_2) & =  &  ( p_1 - p_2  )_\mu \, \mathbb{I}_D \ ,  \\
 L^{(4)}_\mu (p_1 , p_2) & = & \sigma_{\mu\nu} \left( p_1 - p_2 \right)^\nu \ , 
\end{eqnarray}
where $\sigma_{\mu\nu} = {1\over 2} [\gamma_{\mu},\gamma_{\nu}]$.
The transverse operator basis $T^{(i)}_\mu (p_1 , p_2)$ can be found in~\cite{Ball:1980ay}.

\section{Constraints on the Quark-Gluon Vertex}

The global and local symmetries of QCD constraints the full vertex $\Gamma_\mu$ and connect several of the Green's functions of the theory. 
For example, the global symmetries of QCD require that the form factors $\lambda_i$ and $\tau_i$ to be either symmetric or anti-symmetric 
under exchange of the two first momenta -- see, e.g., ref.~\cite{Davydychev} and references therein. 
On the  other hand, gauge symmetry implies non-trivial relations between the QCD Green's functions that can be translated into
Slavnov-Taylor identities~\cite{Slavnov:1972fg,Taylor:1971ff} that play a major role in our understanding of the theory. 
In particular, the transverse part of the quark-gluon vertex is constrained, in the Landau gauge, by the following identity
\begin{multline}
   p^\mu_3 ~ \Gamma_\mu ( p_1, p_2, p_3 ) =   F(p^2_3) ~ \Big [ S^{-1} ( -p_1) \, H( p_1, p_2, p_3 ) ~  \\
    -  ~  \overline H (p_2, p_1, p_3) \, S^{-1}(p_2) \Big ] \ ,
\label{STI}
\end{multline}
where the ghost-dressing function $F(q^2)$ is related to the ghost two-point correlation function as
\begin{equation}
D^{a b} (q^2) = -\, \delta^{ab} \, F(q^2) / q^2
\end{equation}
and $H$ and $\overline H$ are associated to the quark-ghost kernel -- see~\cite{Davydychev} for definitions and discussion.
The quark-ghost kernel can be parametrized in terms of four form factors
\be
  H( p_1, p_2, p_3 ) =  X_0\, \mathbb{I}_D +  X_1 \, \feynslash p_1 +  X_2 \,\feynslash p_2 + X_3 \, \sigma_{\alpha\beta} p^\alpha_1 p^\beta_2\, ,  \nonu
 \overline H( p_2, p_1, p_3 ) = \overline X_0\, \mathbb{I}_D -  \overline X_2\, \feynslash p_1 - \overline X_1\, \feynslash p_2 +
     \overline X_3 \, \sigma_{\alpha\beta} p^\alpha_1 p^\beta_2\,  , \nonu
 \hspace{6mm}
\ee
where $X_i \equiv X_i ( p_1, p_2, p_3 )$ and $ \overline X_i \equiv \overline X_i( p_2, p_1, p_3 )$. 
The above STI  can be solved  for the form factors $\lambda_i (p_1, p_2, p_3)$ that can be written in terms of $A(p^2)$,
$B(p^2)$, $X_i$ and $\overline X_i$ as given in~\cite{Aguilar:2010cn} 
\begin{eqnarray}
\lambda_1 (p_1, p_2, p_3) & =   &
               \frac{F(p^2_3)}{2} \Bigg\{ 
 ~B(p^2_1) \left[X_1 + X_2 \right]   \nonumber  \\ 
 & & \hspace{0.5cm} +  B(p^2_2) \left[ \overline X_1 + \overline X_2   \right]             
               \nonumber \\ 
  & & \hspace{0.5cm} +
               A(p^2_1) \left[ X_0 + \left( p^2_1 - p_1 \cdot p_2 \right) X_3 \right]  \nonumber \\
  & &     \hspace{0.5cm}          
              +  ~ A(p^2_2) \left[ \overline X_0 + \left( p^2_2 - p_1 \cdot p_2 \right) \overline X_3 \right]  
              \Bigg\} \ , \nonumber \\
              \label{EQ:lambda_1} 
 \end{eqnarray}
 \begin{eqnarray}            
\lambda_2 (p_1, p_2, p_3) & = &
               \frac{F(p^2_3)}{2\! \left( p^2_2 - p^2_1 \right) } 
               \Bigg\{ ~B(p^2_1) \left[X_2 - X_1 \right]  \nonumber \\
               & & \hspace{0.5cm} + ~ B(p^2_2) \left[ \overline X_1 - \overline X_2 \right]  \nonumber   \\
               & & \hspace{0.5cm} +~A(p^2_1)\! \left[ \left( p^2_1 + p_1\! \cdot p_2 \right) X_3  - X_0 \right]   \nonumber \\
               & & \hspace{0.5cm} +  ~ A(p^2_2) \left[ \overline X_0 - \left( p^2_2 + p_1\! \cdot p_2 \right) \overline X_3 \right] 
                 \Bigg\} \, , \nonumber \\
                \label{EQ:lambda_2} 
\end{eqnarray} 
\begin{eqnarray} 
\lambda_3 (p_1, p_2, p_3) &  =  &
                   \frac{F(p^2_3)}{p^2_1 - p^2_2} \Bigg\{ ~ B(p^2_1) \, X_0 \nonumber \\
                   & & \hspace{0.5cm} - ~ B(p^2_2) \, \overline X_0   \nonumber  \\ 
                   & & \hspace{0.5cm} + ~ A(p^2_1) \left[ p^2_1 \, X_1 + p_1 \cdot p_2 \ X_2\right] \nonumber \\
                   & & \hspace{0.5cm}  - ~ A(p^2_2) \left[ p^2_2\ \overline X_1 + p_1 \cdot p_2 \ \overline X_2 \right]   
                      \Bigg\}  \ , \nonumber \\
                \label{EQ:lambda_3}  
\end{eqnarray}
\begin{eqnarray} 
\lambda_4 (p_1, p_2, p_3)  & = &  - \frac{F(p^2_3)}{2} \Bigg\{~  B(p^2_1) \, X_3 
  - B(p^2_2) \, \overline X_3 \nonumber \\
& & \hspace{1.3cm} + 
                      A(p^2_1) \, X_2 -  A(p^2_2) \, \overline X_2   \Bigg\}    \ .                                  
\label{EQ:lambda_4} 
\end{eqnarray}
By direct inspection one can show straightforwardly that the solution given in Eqs. (\ref{EQ:lambda_1}) to (\ref{EQ:lambda_4}) for the $\lambda_i$
satisfy the symmetry requirements due to charge conjugation independently of the functions $A$, $B$, $X_i$ and $\overline X_i$. 
This is a particularly important point for modelling the quark-gluon vertex.

\section{The Soft Gluon Limit - an exact relation} 

A great deal of information can be learned from the above expressions by taking its soft gluon limit, i.e. by setting $p_{3}=0$. The
interest on this particular kinematical configuration is that it has been investigated using lattice methods for full QCD with two flavors of quarks. 
In order to avoid the apparent singularity in the expressions for $\lambda_2$ and $\lambda_3$, one can set
$p_1= p $, $p_2 = -p +\delta$, $p_3 = -\delta$, 
expand all quantities in (\ref{EQ:lambda_1})--(\ref{EQ:lambda_4}) up to first order in $\delta$ and, finally set $\delta = 0$.
If one takes into account the all-order results of~\cite{Aguilar:2014lha} which states that, in the soft gluon limit,
$X_0(p^2) = 1$ and $X_1(p^2) = X_2(p^2)$, note the different notation concerning the definitions for $X_1$ and $X_2$,
it follows that
\begin{eqnarray}
\lambda_1 (p^2) & =  & \frac{F(0)}{Z(p^2)} \Bigg\{  1 + 2 \, X_1(p^2) \, M(p^2)  + 2 \, p^2 \, X_3(p^2) \Bigg\}  \, , \nonumber \\
     \label{EqL1soft} \\
\lambda_2 (p^2) & =  & \frac{F(0)}{2 \, Z(p^2)} \Bigg\{ \frac{A'(p^2)}{A(p^2)}  - X_3 (p^2) \nonumber \\
                    & & \hspace{1.5cm}  + ~ \partial_1 X_0 (p^2) 
- \partial_2 X_0 (p^2)  \nonumber \\
                        & & \hspace{1.5cm} + ~ M(p^2) \bigg[ \partial_1 X_1(p^2) + \partial_2 X_2 (p^2)  \nonumber \\
                        & &                                                        \hspace{3cm} - \partial_2 X_1 (p^2) - \partial_1 X_2 (p^2)  \bigg]\Bigg\}  , \nonumber \\
                  \label{EqL2soft}\\
\lambda_3 (p) &= & \frac{F(0)}{Z(p^2)} \Bigg\{  \frac{B'(p^2)}{A(p^2)}  +   X_1 (p^2)  \nonumber \\
                  & & \hspace{1.25cm} +  M(p^2)  \bigg[ \partial_1 X_0 (p^2) - \partial_2 X_0 (p^2) \bigg] \nonumber \\
     & & \hspace{1.25cm} + ~ p^2 \, \bigg[ \partial_1 X_1(p^2) + \partial_2 X_2 (p^2) \nonumber \\
     & & \hspace{2.25cm} - \partial_2 X_1 (p^2) - \partial_1 X_2 (p^2)  \bigg]\Bigg\}  \, , \label{EqL3soft}
\end{eqnarray}
with $\lambda_4  (p^2) = 0$ and 
where $\lambda_i(p^2) = \lambda_i(p, -p, 0)$, 
\begin{displaymath}
A'(p^2) = \frac{d A(p^2) }{ p^2},  \qquad\qquad B'(p^2) = \frac{d B(p^2) }{ d p^2} ,
\end{displaymath}
$X_i (p^2) = X_i(p^2, p^2, 0) = \overline X_i(p^2, p^2, 0) =
\overline X_i (p^2)$ and $\partial_i X_j (p^2)$ is the partial derivative of $X_j(p^2_1, p^2_2, p^2_3)$ with
respect to argument $i$ taken at $p^2_1 = p^2_2 = p^2$ and $p^2_3 = 0$. 
The expressions given in (\ref{EqL1soft}) -- (\ref{EqL3soft}) imply the following exact relation 
\begin{eqnarray}
& & \frac{ Z(p^2)}{F(0)} \Bigg\{  \lambda_1 (p^2) + 4 \, p^2 \, \lambda_2 (p^2) - 2 \, M(p^2) \, \lambda_3 (p^2)\Bigg\} 
    =  \nonumber \\
& &  \hspace{0.4cm}  = 1 - 2 \, M(p^2) \, M'(p^2)  \nonumber \\ 
& &   \hspace{1cm} + ~ 2 \bigg( p^2 - M^2(p^2) \bigg) \times \nonumber \\
& & \hspace{2cm} \times \bigg[ \partial_1 X_0 (p^2) - \partial_2 X_0(p^2) - \frac{ Z'(p^2) }{Z(p^2)} \bigg]
   \label{Eq:exact}
\end{eqnarray}
between the non-vanishing longitudinal form factors. The r.h.s. of Eq. (\ref{Eq:exact}) requires only the knowledge of the quark propagator functions 
$Z(p^2)$ and $M(p^2)$ and of the quark-ghost kernel scalar form factor $X_0$. Further, for on-shell momenta, i.e. when
$p^2 = M^2(p^2)$, the relation between the various form factors is independent of the quark-ghost kernel form factor $X_0$ and requires
only the knowledge of the running quark mass. In principle, the functions $M(p^2)$, $Z(p^2)$ and its derivatives seem to smooth which suggest that
the r.h.s. should not diverge either for infrared or ultraviolet momenta. Our present knowledge of $X_0$ also suggests that this quark-ghost kernel
form factor and its derivatives are also finite. If this is the case, then the l.h.s. of Eq. (\ref{Eq:exact}) should also be always finite, preventing
any divergence of the combination $\lambda_1 (p^2) + 4 \, p^2 \, \lambda_2 (p^2) - 2 \, M(p^2) \, \lambda_3 (p^2)$. 
In other words, if any of the form factors diverges, its divergence has to be compensate by the remaining form factors.

\section{The Lattice Soft Gluon Quark-Gluon Vertex}

\begin{figure}[t] 
   \centering
   \includegraphics[width=2.7in]{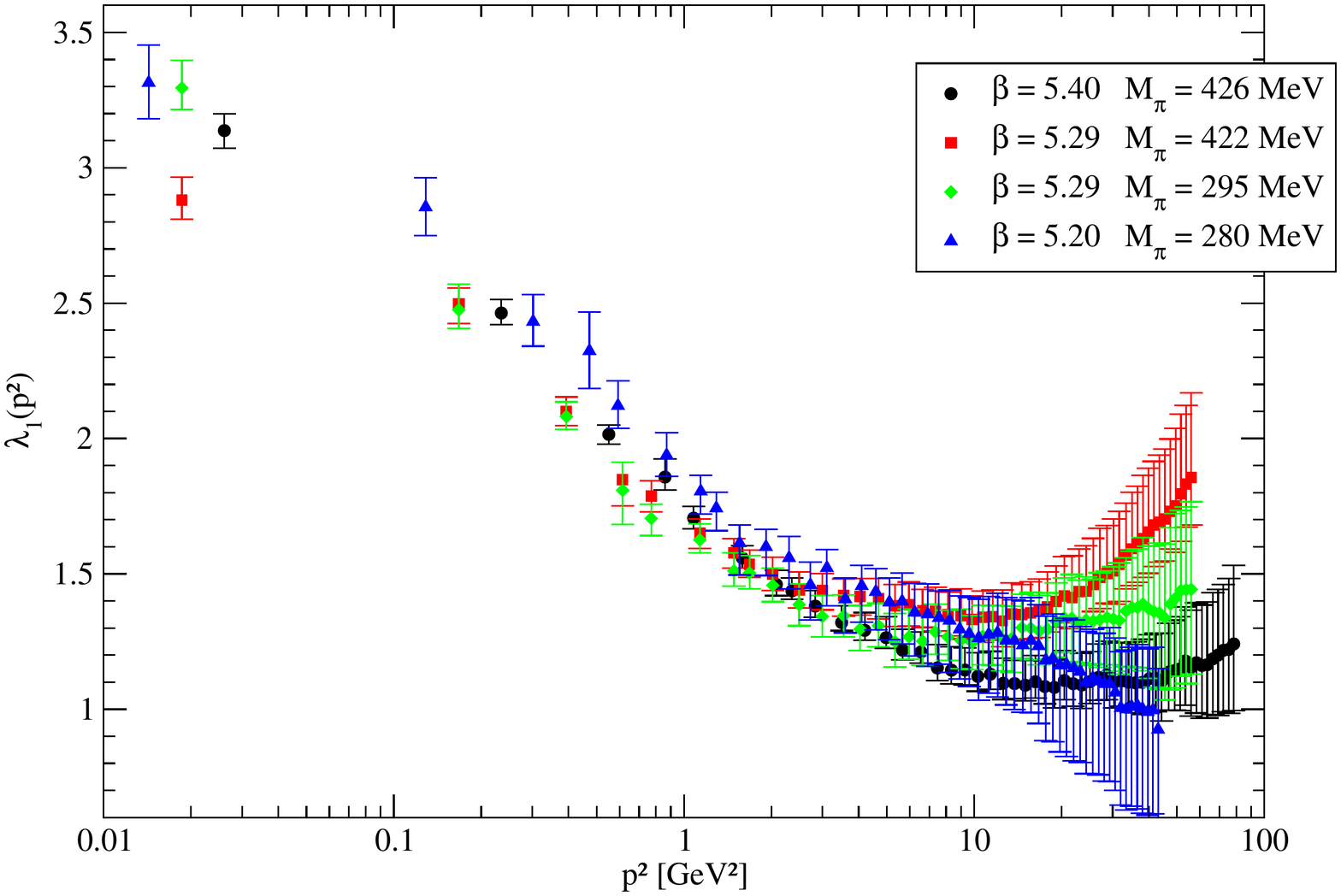} 
   \includegraphics[width=2.7in]{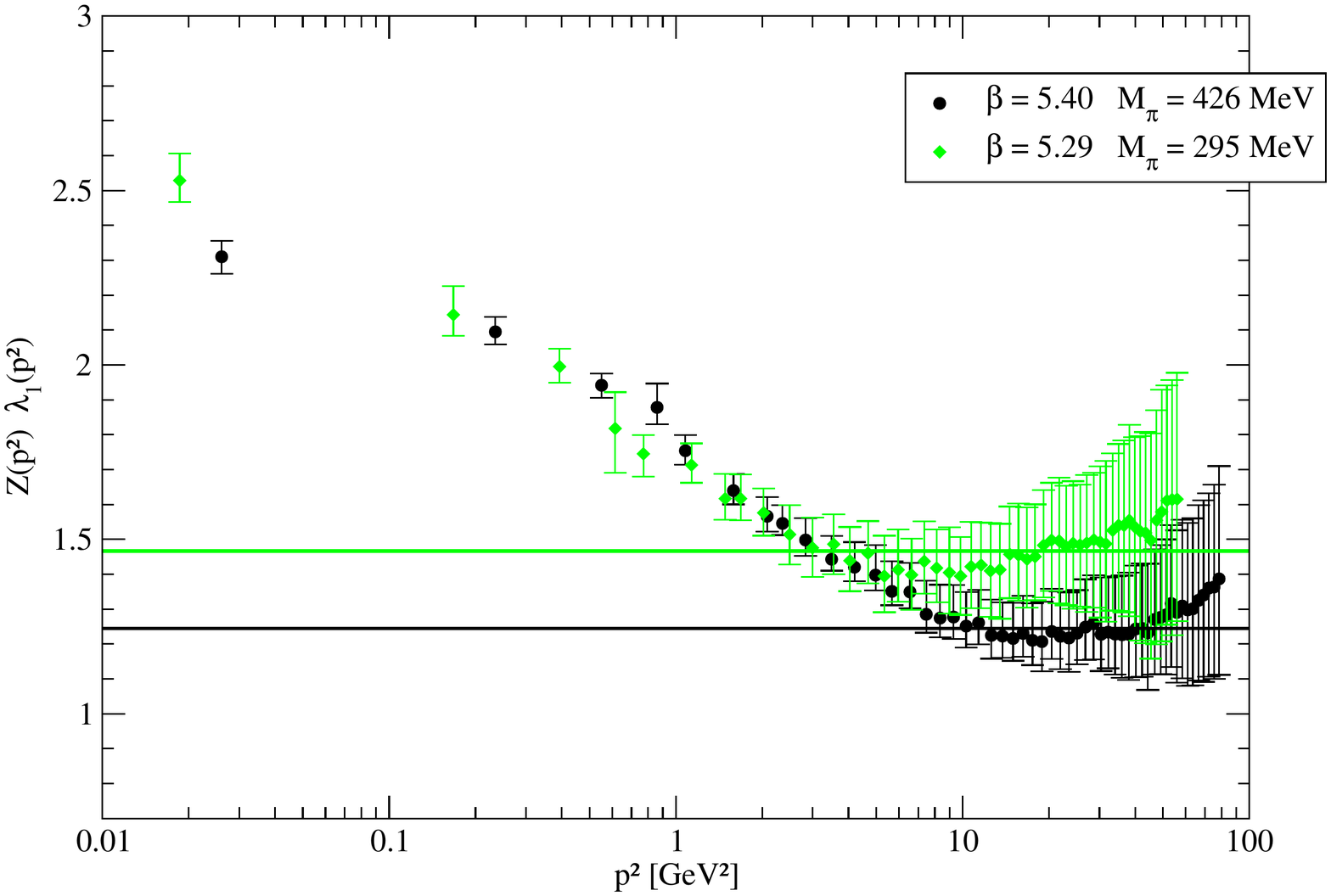}
   \caption{ Tree level corrected lattice data from the full QCD $N_f = 2$ lattice $32^3\times64$ simulations for $\lambda_1(p^2)$ in 
                  the Landau gauge~\cite{Oliveira:2016muq} (top-frame). 
                 $Z(p^2)\,\lambda_1(p^2)$ as a function of $p^2$ (bottom-frame). See text for the explanation of the constant lines.               
                 }
   \label{fig:lambda1Z}
\end{figure}

In~\cite{Oliveira:2016muq} one can find full QCD lattice simulations for the quark propagator and $\lambda_1 (p^2)$
for $N_f = 2$, in the Landau gauge, and for various lattice spacings. For completeness, 
on Fig.~\ref{fig:lambda1Z} top-frame we reproduce the bare lattice results for $\lambda_1 (p^2)$.
In the Landau gauge, at high momenta one expects $\lambda_1(p^2)$ to approach a constant value, up to logarithmic corrections. 
As the top of Fig.~\ref{fig:lambda1Z} shows this is not the case for all the lattice data. In particular, the simulations
performed with $\beta = 5.29$ and a $M_\pi = 422$ MeV together with that with $\beta = 5.20$ and a $M_\pi = 280$ MeV are hardly constant at high momenta.
This is certainly the result of a relative poor estimation of the lattice artefacts. A problem that has to be addressed by the lattice practitioners.
In order to avoid possible contaminations of the lattice artefacts, in the analysis of the lattice data we will not take into account these two sets of results.

The lattice results for $\lambda_1(p^2)$ and $Z(p^2)$ provided in~\cite{Oliveira:2016muq},
combined with Eq. (\ref{EqL1soft}), can be used to estimate
\begin{eqnarray}
& & Z(p^2) \,  \lambda_1(p^2) = \nonumber \\
& & \hspace{.6cm} = F(0) \, \Bigg\{  1 + 2 \, X_1(p^2) \, M(p^2)  - 2 \, p^2 \, X_3(p^2) \Bigg\}  \ . \label{Eq:teste1}
\end{eqnarray}
Note the minus sign relative to Eq. (\ref{EqL1soft}) due to the Wick rotation to the Euclidean space.
On the bottom-frame of Fig.~\ref{fig:lambda1Z} we report the lattice estimates for $Z(p^2) \, \lambda_1(p^2)$. The data for $Z(p^2)$ was interpolated to match the momenta
available for $\lambda_1(p^2)$ and, at high momenta, we set $Z(p^2)$ to a constant matching its higher value. The lattice data reported on~\cite{Oliveira:2016muq}
show a slightly decreasing function $Z(p^2)$ for $p \gtrsim 2.5$ GeV. As discussed in the publication, this seems to be a lattice artefact and
extrapolations to the ``continuum'' suggest that $Z(p^2)$ should, indeed, take a constant value. 

The lattice data reported on the bottom-frame of Fig.~\ref{fig:lambda1Z} show a $Z(p^2) \, \lambda_1(p^2)$ that is constant, within the precision of the lattice
simulation, for $p \gtrsim 2.5$ GeV. This complies with the predictions of perturbation theory which, at tree
level, give $X_1(p^2) = X_3(p^2) = 0$. In principle the logarithmic corrections for $\lambda_1(p^2)$ should show up at high momenta.
However, given the relative large errors on the lattice data for this form factor and given that the lattice data ends at $p  \sim 10$ GeV, it is impossible
to make any statement about the high momenta log behaviour for $\lambda_1(p^2)$.

The observed constant value is an estimation of the bare $F(0)$ value. To get a reference number for
$F(0)$ we fitted the lattice data for $p > 3$ GeV onwards to a constant. For the data referring to the $\beta = 5.40$ and $M_\pi = 426$ MeV simulation, the fit
taking into account the correlations between the momenta gives a bare $F(0) = 1.2444(55)$ with a $\chi^2/d.o.f. = 0.05$, while for the simulation using 
$\beta = 5.29$ and having a $M_\pi = 295$ MeV the correlated fit give a bare $F(0)  = 1.4665(90)$ for $\chi^2/d.o.f = 0.08$. These figures are the constant lines
appearing on the bottom-frame of Fig.~\ref{fig:lambda1Z}. If one takes the smallest lattice momentum for each simulation as reference, it follows that relative to the constant fitted
value $Z(p^2) \lambda_1(p^2)$ increases by a multiplicative factor of 1.86, for the $\beta = 5.40$ and $M_\pi = 426$ MeV simulation, and of 1.72,
for the $\beta = 5.29$ and  $M_\pi = 295$ MeV simulation. Then, from Eq. (\ref{Eq:teste1}) one gets
\begin{displaymath}
   \left[ 2 \, X_1(p^2) \, M(p^2)  - 2 \, p^2 \, X_3(p^2) \right]_{p^2 = 0} = 0.86
\end{displaymath}
for the $\beta = 5.40$ simulation and
\begin{displaymath}
   \left[ 2 \, X_1(p^2) \, M(p^2)  - 2 \, p^2 \, X_3(p^2) \right]_{p^2 = 0} = 0.72  
\end{displaymath}
for the $\beta = 5.29$ simulation.
The finiteness of such a quantity means that either there are no infrared divergences on $X_1(p^2)$ and $p^2 X_3(p^2)$ or they cancel exactly. 
%
%

\begin{figure}[t] 
   \centering
   \includegraphics[width=3.5in]{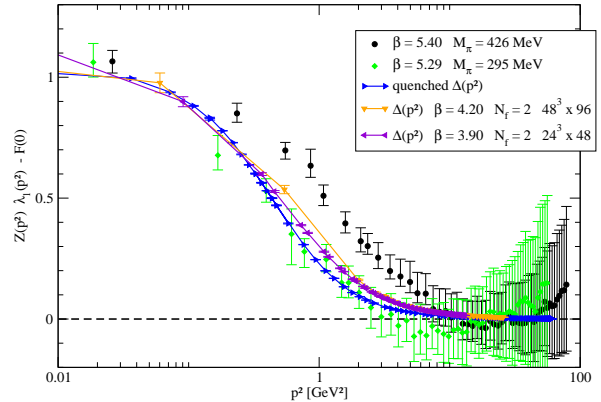}
   \caption{Full QCD $N_f = 2$ lattice $32^3\times64$ simulations for $\lambda_1(p^2)$ for the Landau gauge~\cite{Oliveira:2016muq}
                 together with rescaled Landau gauge gluon propagator for pure Yang-Mills theory~\cite{Duarte:2016iko,Dudal:2018cli}
                 and full QCD~\cite{Ayala:2012pb} for $N_f = 2$.}
   \label{fig:lambda1timesZgluon}
\end{figure}

In what concerns the form factor $\lambda_1 (p^2)$ we observe that $Z(p^2) \, \lambda_1 (p^2) - F(0)$ seems to be proportional to the Landau gauge
lattice propagator. This can be seen on Fig.~\ref{fig:lambda1timesZgluon} where the lattice data is plotted together with the 
rescaled quenched Landau gauge gluon propagator, from a simulation using a $64^4$ lattice and $\beta = 6.0$~\cite{Duarte:2016iko,Dudal:2018cli}, 
and the two full QCD Landau gauge rescaled gluon propagators from simulations using $N_f = 2$ reported in~\cite{Ayala:2012pb}. The good
agreement between the set of data closer to the physical limit is impressive. 
On Sec.~\ref{Sec:final} we are able to link  $\lambda_1(p^2)$ and the gluon propagator relying on a Bethe-Salpeter equation
for the vertex.

The understanding of the remaining  soft gluon form factors $\lambda_2( p^2)$ and $\lambda_3(p^2)$, see Eqs. (\ref{EqL2soft}) and  (\ref{EqL3soft}),
is more cumbersome as their dependence on the quark propagator and quark-ghost kernel form factors is more elaborated.

For $\lambda_2(p^2)$, the estimates performed in~\cite{Aguilar:2014lha} do not provide a very clear picture. 
Indeed, the authors discuss two different calculations. 
In their simplest analysis for $\lambda_2(p^2)$, this form factor is small over the full range of momenta. 
However, by improving the angular dependence, that is taken into account partially, then $\lambda_2(p^2)$
seems to have an infrared divergence. This later result does not seem to comply with our analysis.
The computation of $\lambda_2(p^2)$ within the Curci-Ferrari model~\cite{Pelaez:2015tba} points towards a small contribution from this
form factor that reaches a maximum of $\sim 0.2$ GeV$^{-2}$ at zero momenta or close by. 
The results of one-loop dressed perturbation theory~\cite{Aguilar:2016lbe} also suggest a small $\lambda_2(p^2)$ at small momenta. 
In what concerns lattice estimations of $\lambda_2 (p^2)$, one can find in the literature only results for quenched simulations~\cite{Skullerud:2003qu}. 
The lattice form factor does not agree with the estimates just mentioned for $\lambda_2(p^2)$ and suggests a functional behaviour that is not
necessarily closer to the other estimations. There is an ongoing discussion about the estimation
of the lattice artefacts and a possible contamination of the lattice calculation due to the transverse form factors; 
see e.g. Ref. \cite{Pelaez:2015tba}.

\begin{figure}[t] 
   \centering
   \includegraphics[width=3.5in]{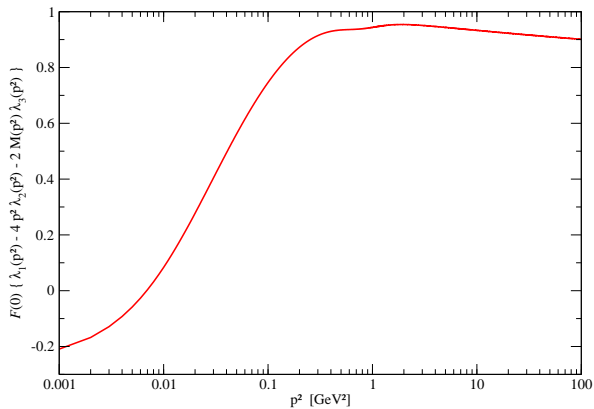}
   \caption{Estimation of the r.h.s. of the Eq. (\ref{Eq:exact}) multiplied by $Z(p^2)$ using the fits provided in~\cite{Rojas:2013tza}. Note the change
                  on the sign for the $\lambda_2(p^2)$ due to the Wick rotation to the Euclidean space.}
   \label{fig:exactrelation}
\end{figure}

For the form factor $\lambda_3(p^2)$ the estimates of \cite{Aguilar:2014lha,Aguilar:2016lbe,Pelaez:2015tba} and the results of quenched lattice
simulations~\cite{Skullerud:2003qu} point towards a sizeable form factor. Given our lack of knowledge of the various terms
in (\ref{EqL3soft}) it is difficult to provide any information on the form factor directly from this equation. 
However, 
after dividing Eq. (\ref{Eq:exact}) by $Z(p^2)$, its r.h.s.  contains only information about the quark propagator and the derivatives
of $X_0$. This relation is a non-trivial constraint on the non-vanishing $\lambda_i (p^2)$. 
Unfortunately, the r.h.s. of the equation requires the knowledge of the derivatives of the mass function and of the quark wave function
that are difficult to evaluate directly from the lattice data. 
One can estimate the r.h.s. of the equation using the fits provided in~\cite{Rojas:2013tza} as shown on Fig.~\ref{fig:exactrelation} 
if one ignores the contribution of the $X_0$ derivatives.
 In the UV region it is essentially flat for momenta $p \gtrsim 1$ GeV and it decreases significantly only for $p \lesssim 300$ MeV. Given
that $\lambda_2(p^2)$ is always small and is multiplied by $p^2$, 
that $\lambda_1(0) \approx 1$, as shown on  Fig.~\ref{fig:lambda1Z},
this can be seen as an indication that $\lambda_3 (p^2)$ takes large values
only for $p \lesssim 300$ MeV, in agreement with the results observed in the  cited articles.

\section{Discussion \label{Sec:final} }

\begin{figure*}[t] 
  \vspace{-.3cm}
   \centering
   \includegraphics[width=4.5in]{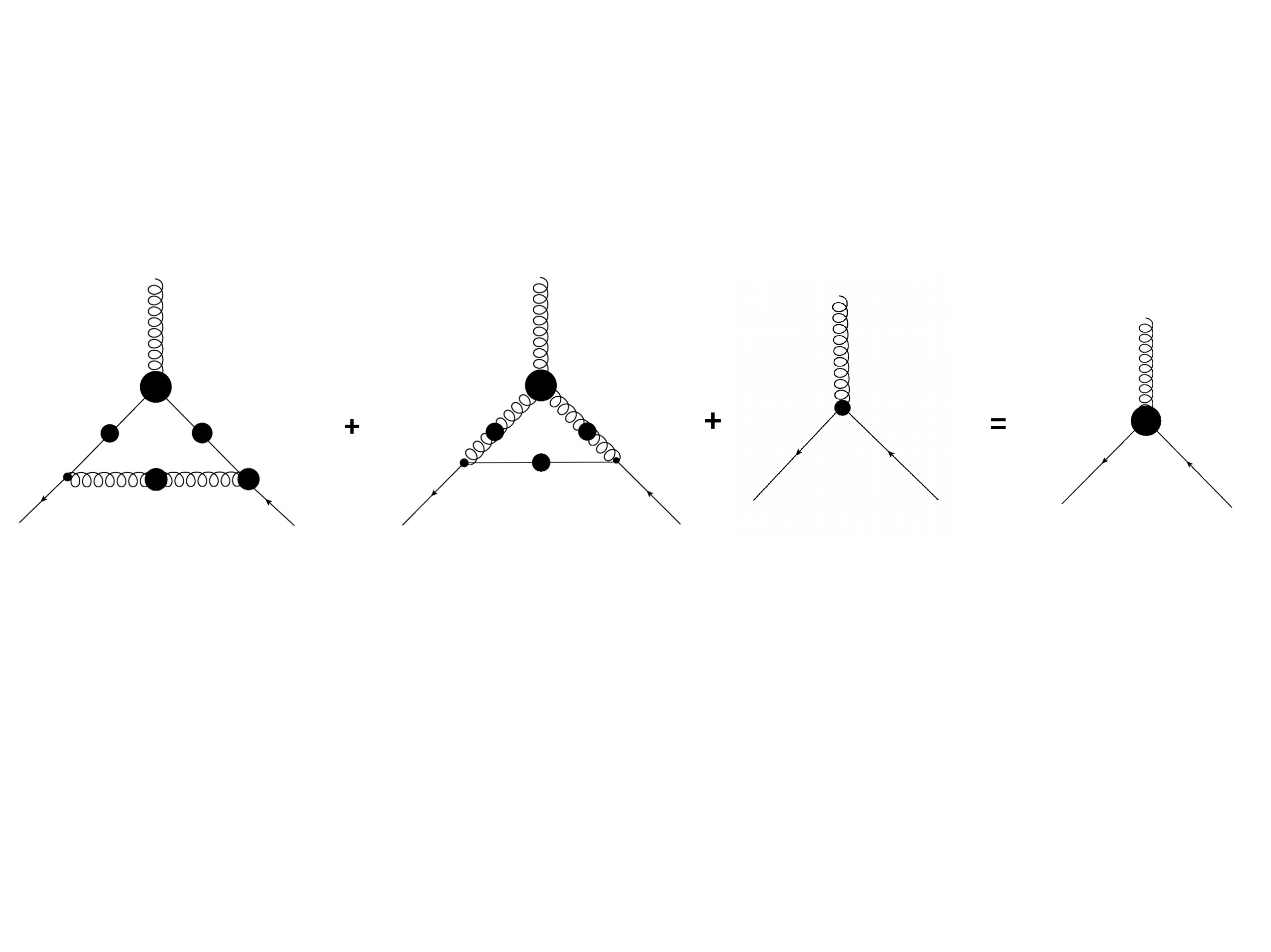} 
    \vspace{-3cm}
   \caption{One-loop dressed Bethe-Salpeter equation for the quark-gluon vertex. The big blobs represent the dressing, and small blobs are bare  vertices.}
   \label{fig:VERTEX1}
\end{figure*}

The observations  reported on the previous section for the soft-gluon limit can be understood looking at the one-loop Bethe-Salpeter equation 
(BSE), in the fully dressed  ladder approximation.
The BSE is represented diagrammatically on Fig. \ref{fig:VERTEX1} where blobs stand for fully dressed quark and gluon propagators,
with the dressed vertices represented by big blobs.
Coming from left to right in the figure the diagrams represent: (i) the abelian like dressed-one gluon exchange ladder kernel contribution;
(ii) the non-abelian like contribution due to the dressed three-gluon vertex;  (iii) the bare quark-gluon vertex. 
These contributions add up to define the self-consistent non-perturbative dressed quark-gluon vertex represented on the right-hand side of the equality. 

The kernel of the BSE represented on Fig.~\ref{fig:VERTEX1} contains a plethora of other types of diagrams as, for example the cross-ladder 
contributions associated to non-planar diagrams. However, as shown in~\cite{thooft,Hornbostel90}, 
in an expansion in the number of colors, the non-planar diagrams and the diagrams not represented are subleading. 
In addition, the numerical solution of the BSE~\cite{Nogueira:2017pmj} for a bosonic model with color degrees of freedom, 
using a kernel of the BSE for the two-boson bound state amplitude that contains the ladder and the lowest non-planar cross-ladder contribution, 
shows that the cross-ladder kernel contributes less than a few-percent to the binding energy and to the elastic electromagnetic form factor.

In principle the driving contribution in the diagrams represented on Fig.~\ref{fig:VERTEX1} is the diagram with higher number of full vertices, i.e. the
abelian-like term of the left, and its contribution is proportional to
\begin{multline}
    \int \frac{d^4 q}{( 2 \, \pi )^4}    
     \gamma_\beta \,  S(p_2 - q )  \,  \Gamma_\mu (-p_2-p_3+ q, p_2 - q, p_3)  \times 
\\    
    \times S(-p_1-q)  \, \Gamma_\alpha (p_1, -p_1-q, q) 
    ~  \left( g^{\alpha\beta} - \frac{q^\alpha q^\beta}{q^2} \right) \, \Delta (q^2)  , \label{Eq:IRanalysis}
\end{multline}
where $p_2$ is the incoming quark momenta, $-p_1$ the outgoing quark momenta and $p_3$ the incoming gluon momenta. 
Note that the second diagram from left on Fig.~\ref{fig:VERTEX1} is proportional to the three-gluon non-abelian vertex, whose coupling is 
proportional to the momentum of the gluons attached. Due to the presence of the gluon momenta, in the soft gluon limit we expect this contribution to be subleading.
In the soft gluon limit
$p_1 = p_2 = p$, $p_3 = 0$ and $\Gamma_\mu (-p_2-p_3+ q, p_2 - q, p_3) $ is replaced by its soft gluon version that is dominated either by
$\lambda_1 ( (p_2 - q)^2)$ or $\lambda_3 ( (p_2 - q)^2)$. In both cases the form factors reach their maximum values in the deep infrared region and, therefore,
the major contribution due to the momentum integration in Eq. (\ref{Eq:IRanalysis}) occurs for $p_2 - q \approx 0$. Then, in the soft gluon limit, the BSE
predicts a contribution to the vertex that is proportional to $\Delta (p^2)$ and that adds to the tree level vertex, as observed on Fig.~\ref{fig:lambda1timesZgluon}.
This analysis suggests that we may write
\begin{equation}
\lambda_1(p^ 2)\, Z(p^2)\approx a_L\,+\, b_L\,\Lambda_{QCD}^2\, \Delta(p^2) \, , \label{approxd}
\end{equation}
where the ultraviolet constant term is represented by $a_L$ and the dimensionless quantities $a_L$ and $b_L$ should be of order $\mathcal{O}(1)$. The previous
discussion when building Fig. ~\ref{fig:lambda1timesZgluon} give a $a_L = F(0) \approx 1.24$ for the 
$\beta = 5.40$ ($M_\pi = 426$ MeV) data and a $a_L = F(0) \approx 1.47$ for $\beta = 5.29$ ($M_\pi = 295$ MeV) data. The quenched gluon propagator
in Fig.~\ref{fig:lambda1timesZgluon} was renormalized in the MOM-scheme at $\mu = 3$ GeV and it was rescaled by a factor of $\sim 0.12$ GeV$^2$ to produce
the figure. The full QCD gluon propagator for the $\beta = 4.20$ simulation was also renormalized in the MOM-scheme at $\mu = 4.3$ GeV and
rescaled by a factor of $\sim 0.19$ GeV$^2$ to produce the figure. From its definition we have that $b_L = c \, F(0)$, where $c$ is
a dimensionless factor that corrects for the momentum integration. For a $\Lambda_{QCD} = 0.3$ GeV and taking into account the above quoted values for
$F(0)$, the prediction being that $b_L \, \Lambda_{QCD}^2\approx c ( 0.11 - 0.13) $ GeV$^2$ and a $c \sim 1$,  in agreement with the numbers discussed.
Note that both the propagator and the constant $c$ are not independent of the renormalization scale and, therefore, the quoted figures should be
read as orders of magnitude. Note also that the above picture applies to the data closer to the physical limit. Indeed the form factor depends on the 
quark (pion) mass in such a way that for larger values of the mass, the decrease of $Z(p^2) \, \lambda_1(p^2)$ or $\lambda_1 (p^2)$
is slower and, therefore, the above argument does not apply for sufficiently heavier quark masses.

Despite the relative large statistical and lattice artefacts observed on the lattice data, our empirical analysis 
based on Eqs. (\ref{Eq:teste1}) and  (\ref{approxd}) supports the factorization 
of $\lambda_1(p^ 2)\, Z(p^2)$ in such a way that
\begin{multline}
  F(0) \bigg( 2 X_1(p^2) M(p^2)  - 2 p^2 X_3(p^2) \bigg) \\ \approx  a^\prime_L\,+\, b_L\,\Lambda_{QCD}^2\, \Delta(p^2) \, ,
\end{multline}
with $a^\prime_L=a_L-F(0) \approx 0$. It is the gluon propagator that dominates the relevant momentum behaviour of the product $\lambda_1(p^ 2)\, Z(p^2)$
at infrared scales.
Furthermore, this results evidences the ladder dominance, as already suggested by the rainbow-ladder approximation of the Dyson-Schwinger equation for the 
quark propagator and used in many phenomenological applications of continuum methods to treat  strong QCD to study mesons and baryons properties \cite{CloPPNP14}. 
If one assumes that the product $X_1(p^2) M(p^2)$ carries the dominant momentum behavior, then 
$X_1(p^2) M(p^2)\propto \Delta(p^2)$. Such a feature is important in building a quark-gluon vertex model having 
dependence not only on the gluon momentum but also on the quark momentum. Eventually such factorization can be extended to the
other components of the quark gluon vertex in the soft gluon limit.


\section{Summary} 

The longitudinal component of the quark-gluon vertex in the soft gluon limit 
is studied combining information from Lattice QCD simulations and a Slavnov-Taylor identity.
In the soft gluon limit, the Slavnov-Taylor identity allows to write the longitudinal form factors
in terms of the quark-ghost kernel and the quark propagator functions. 
Further, the STI implies, in the soft gluon limit,
an exact relation that combines linearly $\lambda_1$, $\lambda_2$ and $\lambda_3$ with
differences between derivatives of $X_0$, the quark self energy and the quark wave function. 

From full QCD lattice simulation data for $\lambda_1(p^2)$ in the Landau gauge we are able to estimate 
some of the quark-ghost kernel form factors. The data also allows to explore the remaining non-vanishing 
quark-gluon vertex functions $\lambda_2(p^2)$ and $\lambda_3(p^2)$. We found that the provided estimations are 
within typical values found in the literature.

In particular for $\lambda_1 (p^2)$ the analysis of the recent full QCD Landau gauge lattice simulations reveals an empirical relation 
linking $\lambda_1(p^2)$ with the gluon propagator $\Delta (p^2)$. This relation seems to hold for the momenta accessed on the lattice
simulations and within the statistical errors of the lattice data. 
The relation between $\lambda_1(p^2)$ and $\Delta (p^2)$ is found in the solution of a fully dressed one-loop Bethe-Salpeter equation
for the quark-gluon vertex.
The analysis of the BSE suggests that the factorisation holds for other components of the vertex in the soft gluon limit and, eventually, also for
its transverse components. Such challenging investigation is left for future work.
The connection between the $\lambda_i$ and the gluon propagator can contribute to model the quark-gluon vertex
including not only the dependence on the gluon momentum but also the quark momentum.

\section*{Acknowledges}

This work was partly supported by the Funda\c c\~ao de Amparo \`a Pesquisa do Estado de S\~ao Paulo [FAPESP grant no. 17/05660-0],
Conselho Nacional de Desenvolvimento Cient\'ifico e Tecnol\'ogico [CNPq grant 308486/\-2015-3] and 
Coordena\c c\~ao de Aperfei\c coamento de Pessoal de N\'ivel Superior (CAPES) of Brazil.  This work is a part of the project INCT-FNA Proc.  No. 464898/2014-5.
O. Oliveira acknowledge support from CAPES  process 88887.156738/2017-00 and FAPESP grant number 2017/01142-4.




\end{document}